\documentclass[preprint2]{aastex}

\newcommand {\asp}{\mbox{$.\!\!^{\prime\prime}$}}

\slugcomment{To appear in The Astrophysical Journal (Letters)}

\shorttitle{Protoplanetary disk in CB\,26}
\shortauthors{Launhardt and Sargent}

\begin{document}

\title{A young protoplanetary disk in the Bok globule CB\,26?}

\author{R. Launhardt\altaffilmark{} and A. I. Sargent\altaffilmark{}}
\affil{Division of Physics, Mathematics, and Astronomy, California Institute of Technology, 
    Pasadena, CA 91125}
\email{rl@astro.caltech.edu}


\begin{abstract}

We present sub-arcsecond resolution millimeter-wave images of a 
circumstellar disk in the Bok globule CB\,26. 
The presence of an edge-on disk is confirmed by the dust continuum morphology 
and the velocity field of $^{13}$CO emission, 
which displays a Keplerian rotation pattern about an axis 
perpendicular to the long axis of the dust emission. 
We deduce a mass $\sim$0.3\,M$_{\odot}$\ for the obscured central star. 
The disk is optically thick at mm wavelengths inside 120\,AU, 
has a symmetric 20\degr\ warp beyond 120\,AU, an outer radius of 
$\sim 200$\,AU, and a mass of at least 0.1\,M$_{\odot}$.
We suggest that the CB\,26 system is in an intermediate stage between 
deeply embedded protostellar accretion disks and the more evolved, perhaps 
protoplanetary, disks around T Tauri stars.

\end{abstract}




\section{Introduction} \label{sec_intro}

CB\,26 (L\,1439) is a small cometary-shaped Bok globule 
located $\sim$10\degr\ north of the Taurus/Auriga dark cloud. 
Launhardt \& Henning (1997) suggested a distance of 300\,pc. 
From a re-examination of the larger-scale velocity structure 
(Ungerechts \& Thaddeus 1987), we believe it is 
part of the Taurus/Auriga complex at 140\,pc. 
Single-dish submm images at 10\arcsec\ resolution show strong 
unresolved dust continuum emission at the south-west rim of the globule, 
surrounded by a thin asymmetric envelope with a well-ordered 
magnetic field directed along PA\,$\sim$\,25\degr\ 
(Henning et al. 2001).
A small bipolar near-infrared (NIR) nebula bisected by a dark extinction 
lane is associated with the millimeter emission. 
The sub-arcsecond resolution NIR polarization pattern 
is consistent with the presence of an almost 
edge-on circumstellar disk elongated along the extinction lane 
at PA\,$\sim$\,60\degr\ and a thin envelope 
(Stecklum et al. 2001).
The spectral energy distribution 
indicates a low-mass, $\ge$\,0.5\,L$_{\odot}$\ 
pre-main sequence (PMS) star surrounded by a disk and an envelope 
(Stecklum et al. 2001), i.e., 
a Class\,I object (Adams, Lada, \& Shu 1987; Kenyon \& Hartmann 1995).


\section{Observations and data reduction} \label{sec_obs}

CB\,26 was observed with the Owens Valley Radio Observatory (OVRO) 
millimeter-wave array between January and December 2000. 
Three configurations of the six 10.4\,m antennas provided baselines 
in the range 
6\,--\,180\,k$\lambda$\ at 2.7\,mm (110\,GHz) and 
12\,--\,400\,k$\lambda$\ at 1.3\,mm (232\,GHz).
Average SSB system temperatures of the SIS receivers were 
300\,--\,400\,K at 110\,GHz and 300\,--\,600\,K at 236\,GHz.
The digital correlator was centered on the 
$^{13}$CO(1--0) line at 110.2\,GHz, adopting the systemic velocity 
of CB\,26, $v_{\rm LSR} = 5.5$\,km\,s$^{-1}$; 
spectral resolution and bandwidth were 0.17\,km\,s$^{-1}$\ and 
5\,km\,s$^{-1}$, respectively.
Continuum emission at 1.3 and 2.7\,mm 
was observed simultaneously in 2\,GHz-wide bands, 
except in the highest-resolution configuration, 
where the 4\,GHz wide-band capability of new 1\,mm receivers 
was used. 
Amplitude and phase calibration relied on frequent observations 
of a nearby quasar, resulting in absolute position uncertainty 
of 0\asp2. 
Flux densities are based on observations of Uranus and Neptune, 
with resulting uncertainties of 20\%. 
The raw data were calibrated and edited using the MMA software package 
(Scoville et al. 1993). 
Mapping and data analysis used the MIRIAD toolbox 
(Sault, Theuben, \& Wright 1995).
We also applied maximum entropy deconvolution (MEM) to the data 
and obtained images which are very similar to those obtained by 
standard cleaning.


\section{Results} \label{sec_res}

\begin{figure}
\plotone{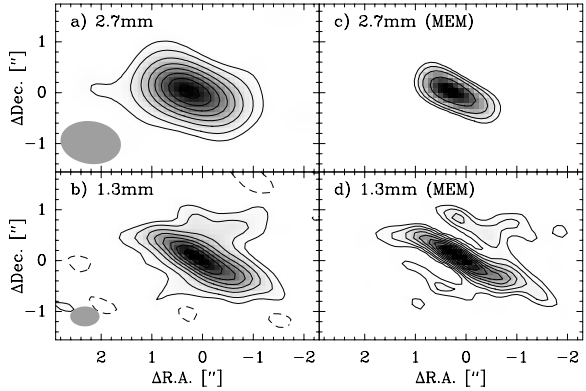}
\caption{\label{fig_mmcres}
   Dust continuum emission from CB\,26 
   (R.A.=04:59:50.74, DEC=+52:04:43.8, J2000):
   1a) and b) Cleaned and restored 2.7 and 1.3mm maps. 
   Contour levels are at -2, 2, 4, 7\ $\ldots$\ 22\,$\times$\,1$\sigma$\,rms 
   where $\sigma=0.7$\,mJy/beam at 2.7mm and 1.3\,mJy/beam at 1.3\,mm. 
   Beam sizes are shown as grey ovals (1\asp2$\times$0\asp84 
   at 2.7mm and 0\asp58$\times$0\asp39 at 1.3mm). 
   c) and d) Deconvolved maps derived using the maximum entropy 
   algorithm (MEM).}
\end{figure}
 
\begin{figure}
\plotone{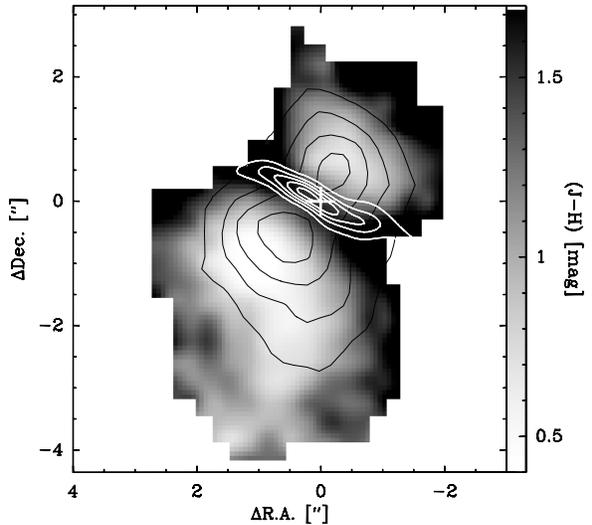}
\caption{\label{fig_jmh}
   MEM image of the 1.3\,mm dust emission (solid white contours at 
   4, 11, 18, 25, and 32\,mJy/arcsec$^2$) 
   overlaid on a J\,$-$\,H color map of the bipolar NIR reflection nebula. 
   Black contours show the nebula K-band emission. 
   The white cross denotes the presumed location of the 
   illuminating star (from Stecklum et al. 2001).} 
\end{figure}

\begin{figure}[thb]
\plotone{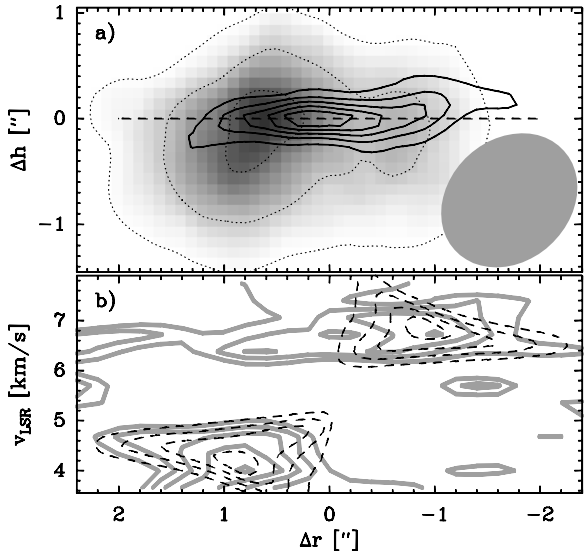}
\caption{\label{fig_cores}
   a) Integrated $^{13}$CO(1--0) emission from CB\,26 in 
   grey-scale (dotted contours at 8, 13.5, 19\,K\,km\,s$^{-1}$) 
   with 1.3\,mm dust continuum contours overlaid. 
   A dashed line represents the plane of the disk. 
   The CO beam size is shown as grey ellipse.~
   b) Position-velocity diagram along the disk major axis. 
      Thick contours at 27, 45, 63, 81, 99\%\ of the maximum 
      intensity show the observed velocity field. 
      The expected emission from a Keplerian disk around a 
      0.35\,M$_{\odot}$\ star is represented by dashed contours 
      (see text).}
\end{figure}

In Fig. \ref{fig_mmcres}, our dust continuum 
images of CB\,26 show a source elongated along PA\,=\,$60\pm 5$\degr.
At 2.7\,mm, the projected length derived from the distribution of 
clean components is 230$\pm$30\,AU. 
The minor axis is unresolved and 
there is no hint of an envelope. 
By contrast, the 1.3\,mm images show a narrow 
central structure of length 400$\pm$40\,AU 
and a small envelope of size 220\,AU\,$\times$\,280\,AU. 
The scale height $h$\ of the central structure remains unresolved, 
i.e., $h$\,$\le$\,20\,AU. 
Beyond radii of 120\,AU, an almost symmetric 20\degr\ warp is 
apparent in Fig. \ref{fig_mmcres}d.
Total continuum flux densities are 22$\pm$4\,mJy and 
150$\pm$30\,mJy at 2.7 and 1.3\,mm, respectively. 
Thus, we recover completely the 1.3\,mm flux of the 
unresolved component as measured with the IRAM 30-m telescope, 160\,mJy 
(Launhardt \& Henning 1997). 
From our maps, we calculate that about 50\,mJy
arises in the envelope and 100\,mJy in the narrow central structure.  
Of this 100\,mJy, about 75\,mJy derives from the inner 230\,AU 
where the 3\,mm emission arises.

In Fig. \ref{fig_jmh}, the 1.3\,mm continuum MEM image is overlaid on 
a map of the NIR reflection nebula. 
Within the NIR astrometric uncertainties, 0\asp5, the dust 
emission coincides with the extinction lane 
and with the location and orientation of the 
disk postulated by Stecklum et al. (2001). 
%
Our kinematic data support the disk hypothesis.
As illustrated in Fig. \ref{fig_cores}a, strong 
$^{13}$CO(1--0) emission ($T_{\rm R}^{\ast}\sim 20$\,K) 
arises mainly from the outer parts of the disk, 
suggesting optically thick emission.
In Fig. \ref{fig_cores}b, the
$^{13}$CO line is clearly double-peaked, with blue-shifted emission 
concentrated to the north-east and red-shifted to the south-west. 
The velocity structure indicates a rotation axis 
perpendicular to the projected plane of the disk.
At the systemic velocity, $v_{\rm LSR}=5.5$\,km\,s$^{-1}$, 
gas emission appears to be self-absorbed, probably due to the extended 
outer envelope.


\section{Discussion} \label{sec_dis} 

Our millimeter interferometric observations of thermal dust continuum 
and $^{13}$CO line emission from CB\,26 confirm the presence 
of an edge-on circumstellar disk as 
postulated by Stecklum et al. (2001).  
The disk has a diameter of about 400\,AU and an as yet 
unexplained warp beyond radii of 120\,AU.  
This could be due to an 
undetected wide companion star with an inclined orbit. 

%
The continuum fluxes lead to an average 1.3-3\,mm dust emission 
spectral index $\alpha$\ for the inner $R$\,$<$\,115\,AU 
of the disk of 1.7$\pm$0.2, where flux $S_{\nu}\propto \nu^{\alpha}$. 
Since $\alpha$\,=\,2 for a blackbody in the Rayleigh-Jeans limit, 
emission from the inner disk must be very optically thick, 
making it difficult to constrain the mass. 
Beyond 120\,AU, no 3\,mm emission is detectable and 
$\alpha$\,$\ge$\,2.7$\pm$0.3, suggesting optically thin emission.
The transition between optically thick and optically thin 
emission occurs approximately at the radius where a warp appears, 
$\sim$\,120\,AU. 
Thus the outer disk may be disturbed and inflated, 
and flared T Tauri star (TTS) disk models 
may no longer apply. 
Nevertheless we derived a lower limit to the disk mass 
by comparing the 1.3 and 2.7\,mm 
continuum emission to spatially resolved radiative transfer calculations 
of different heuristic 'typical' TTS disk models characterized by 
temperature profiles, 
$T_{\rm d} \propto r^{-q}$, with $q$\,=\,0.5\,--\,0.75, 
surface density profiles, 
$\Sigma\propto r^{-p}$, with $p$\,=\,1.0\,--\,1.5, 
and disk flaring, 
$H\propto r^{z}$, with $z$\,=\,0\,--\,1.5 
(e.g., Adams et al. 1987; Kenyon \& Hartmann 1987). 
An interstellar gas-to-dust mass ratio of 100 and typical disk 
dust opacity $\kappa_{1.3}$\,=\,2\,cm$^2$\,g$^{-1}$\ with 
$\kappa\propto \nu^{\beta}$\ and $\beta=1$\ 
were assumed (cf. Beckwith et al. 1990). 
All models suggest disk masses of at least 0.1\,M$_{\odot}$. 
Due to the high optical depths of the $^{13}$CO line, 
self-absorption, and possible depletion (cf. Thi et al. 2001), 
an independent mass estimate from the molecular line 
emission is unobtainable. 

%
The envelope detected in the 1.3\,mm images (Figs. \ref{fig_mmcres}b, 
\ref{fig_mmcres}d) 
is probably associated with an outflow or disk wind.  
Assuming optically thin emission at temperature 30\,K and 
$\kappa_{1.3} = 1\pm 0.5$\,cm$^2$\,g$^{-1}$, a fairly typical 
value for dense protostellar cores (cf. Ossenkopf \& Henning 1994), 
we derive $M_{\rm H}{\rm (env)}=0.01\pm0.003$\,M$_{\odot}$.  
For a cylindrical morphology 
(cf. Figs. \ref{fig_mmcres}b, \ref{fig_mmcres}d, and Sect. \ref{sec_res}), 
we infer an average density $\langle n_{\rm H}\rangle\sim 3\,10^8$\,cm$^{-3}$, 
column density  
$N_{\rm H}\sim 5\,10^{23}$\,cm$^{-2}$, and corresponding visual extinction 
$A_{\rm V}\sim 300$\,mag, implying that 
back-warming from the envelope has to be considered 
in the energy balance of the disk (cf. Butner, Natta, \& Evans 1994). 
The total mass of the more extended ($\sim 3000$\,AU diameter) 
asymmetric envelope seen in single-dish maps 
is $0.1\pm0.05$\,M$_{\odot}$\ (Henning et al. 2001). 
This may be a remnant of the globule core from which the system formed. 

%
The velocity structure of $^{13}$CO can be modeled by Keplerian rotation
when the high optical depth of the inner disk and 
self-absorption due to an extended envelope are taken into account, 
yielding a mass of the central star 
of 0.35$\pm$0.1\,M$_{\odot}$\ (cf. Fig. \ref{fig_cores}b).
However, since the disk mass ($\ge$\,0.1\,M$_{\odot}$) is 
comparable to the inferred mass of the central star,  
self-gravitation may be important in the outer disk 
(cf. Bertin \& Lodato 1999). 
The outer rotation curve would then be flatter than in the pure Keplerian 
case, and the mass of the central star concomitantly lower, 
perhaps only 0.25\,M$_{\odot}$.
The $^{13}$CO line strength suggests a gas-rich disk with 
kinetic temperature $\ge$\,20\,K outside $R\sim 100$\,AU.
An upper limit to the turbulent velocity dispersion 
in the disk is 0.3\,km\,s$^{-1}$. 

%
PMS evolutionary tracks indicate 
an age of $7\pm 2\times 10^5$\,yr for a 0.3\,M$_{\odot}$\ star with 
$L=0.5$\,L$_{\odot}$\ (D'Antona \& Mazzitelli 1994). 
Due to the anisotropic radiation field caused by the optically thick 
edge-on disk, 0.5\,L$_{\odot}$\ may be only a lower limit to 
the bolometric luminosity of the central star
(see Men'shchikov \& Henning 1999). 
But, for such a massive disk, the intrinsic viscous luminosity
could also contribute.
Overall, the properties of the CB\,26 system are consistent with its being 
younger than classical TTSs (ages $10^6--10^7$\,yr). 
There is no dense, centrally peaked cloud core with evidence of 
collapse, and no prominent molecular outflow, indicating that the main 
accretion phase, assumed to last a few times $10^4$\,yr 
(e.g., Andr\'e, Ward-Thompson, \& Barsony 2000), 
has ended.
However, the disk mass is close to the theoretical gravitationally 
unstable limit, 0.3\,M$_{\ast}$, so that 
the envelope may still be accreting at low rates onto the disk
(see Hollenbach, Yorke, \& Johnstone 2000, and references therein).  
Indeed, Fig. \ref{fig_cores}b shows some 'forbidden' red-shifted emission 
from the blue side of the disk which may be due to infall. 

%
We conclude that the CB\,26 disk surrounds a PMS M-type star 
with age $\sim$\,$10^5$\,yr. This represents an 
intermediate stage between such deeply embedded protostellar accretion disks 
as L\,1551\,IRS5 (e.g., Butner et al. 1994), 
and the more evolved, perhaps protoplanetary, disks around classical TTSs, 
which may be the reservoirs for material for planet 
formation.


\acknowledgments

The Owens Valley millimeter-wave array is supported by NSF grant 
AST 9981546. 
Funding from NASA's {\it Origins of Solar Systems} 
program (through grant NAG5-9530) is gratefully acknowledged.
Research at Owens Valley on the formation of young stars and planets is 
also supported by the {\it Norris Planetary Origins Project}. 
We benefited from discussions with B. Stecklum, L. Hartmann, and H. Butner. 


\end{document}